\documentclass[a4paper,10pt]{article}
\usepackage[utf8x]{inputenc}

\title{Gauge Theory on a Four Sphere}
\author{Mir Hameeda \\ Degree Collage Boys Baramulla \\
 Kashmir,  India}

\begin{document}

\maketitle
 
\begin{abstract}
In this paper we will analyse the quantization of a gauge theory on 
a four sphere. This will be done by  mode expanding
 all the fields 
in the theory in terms of harmonic modes. 
 We will also analyse the BRST symmetry 
of this theory.
\end{abstract}

\section{Introduction}
Scalar field theory is quantized by imposing a 
canonical quantization relations 
or by summing over all field configurations.    
If we try to impose the canonical quantization 
relations on a theory with gauge degrees of freedom, 
we will get constraints; and if we try to sum over the 
field configurations, we will get divergence in the 
partition function. One way to deal with this problem is the
Wheeler-DeWitt approach \cite{wd}-\cite{wd1}.
Another way this problem can be dealt with is by fixing a gauge.   
 This is achieved at a quantum level by adding a gauge fixing 
term and a ghost term. The resultant theory thus obtained has 
negative norm states called the ghost states. These states can 
be removed by using a symmetry called the BRST 
symmetry \cite{1a}-\cite{2a}.  
The BRST symmetry been studied for various theories with a gauge symmetry
 \cite{7a}-\cite{8a}.  In the BRST formalism the effective Lagrangian is given by a sum of 
the original classical Lagrangian, the gauge fixing term and the ghost term. 
The charge corresponding to the invariance of the effective Lagrangian under BRST symmetry is nilpotent. It is thus  used to
factor out the physical states.  In this paper we will analyse
 a gauge theory on a four sphere and apply the BRST formalism to it.

\section{Associated Legendre Function}
In this section we will review some properties of associated Legendre function. These properties will be used to construct various spherical harmonics.
In general an associated Legendre function $P^{-\mu}_{\nu}(x)$ is given by
\begin{equation}
P^{-\mu}_{\nu}(x) = \frac{1}{\Gamma(1+\mu)}\left(\frac{1-x}{1+x}\right)^{\frac{\mu}{2}} F(-\nu, \nu+1, \mu+1, \frac{1-x}{2}).
\end{equation}
Now $\Gamma(1+\mu) $ is the Gamma function and  $F(\nu, \nu+1, \mu+1, \frac{1-x}{2})$ is the hypergeometric function. The hypergeometric function  $F(a,b,c,x)$ in general is given by
 \begin{equation}
  F(a,b,c,x) = 1 + \frac{ab}{c}x + \frac{a(a+1)b(b+1)}{2! c(c+1)} x^2 + \cdots.
 \end{equation}
We will need the lowering and raising operators for  $\nu$ to find relations between various spherical harmonics.

The lowering operator for $\nu$ is given by
\begin{equation}
 \left(  (1-x^2)\frac{d}{dx} +\nu x\right)P^{-\mu}_{\nu}(x) = (\nu-\mu)P_{\nu-1}^{-\mu}(x).
\end{equation}
The raising operator for $\nu$ is given by
\begin{equation}
 \left(  (1-x^2)\frac{d}{dx} -(\nu+1) x\right)P^{-\mu}_{\nu}(x) = -(\nu+\mu+1)P_{\nu+1}^{-\mu}(x).
\end{equation}
It will be useful to  define $D_m$ as follows:
\begin{equation}
 D_m = \frac{d}{d\chi} + m\cot\chi.
\end{equation}
Then as
\begin{eqnarray}
  \left[ \frac{d}{d\chi} + m\cot\chi \right] (\sin\chi)^n f(\chi) = && \nonumber \\
  \sin^n\chi \left[ \frac{d}{d\chi} + (m+n)\cot\chi \right] f(\chi),&&
\end{eqnarray}
 we can write
\begin{equation}
 D_m \sin^n \chi f(\chi) = \sin^n \chi D_{m+n} f(\chi).
\end{equation}
We also have 
\begin{equation}
 -\sin \chi D_n = \left[ (1- \cos\chi^2)\frac{d}{d\cos\chi} - n \cos\chi \right]. 
\end{equation}

\section{ Spherical Harmonics}
In this section we will  review spherical harmonics 
on $S^4$ \cite{air}-\cite{ir1}. We will first see how  we can construct  
scalar spherical harmonics on $S^n$, if we are given the
 scalar spherical harmonics on $S^{n-1}$. 
Then we will explicitly construct the scalar spherical harmonics on $S^4$.
 The metric on $S^n$ can be written as follows:
\begin{equation}
 ds^2 = d\chi^2 + \sin\chi^2 ds_{n-1}.
\end{equation}
Here $ds_{n-1}$ is the metric on $S^{n-1}$.
We can define a function $^n P^l_L$ as
\begin{equation}
 {^n P^l_L} (\chi) =  c_n (\sin \chi)^m P^{-l+m}_{L-m}(\cos \chi),
\end{equation}
where $ m = (2-n)/2$ and 
 $ c_n $ is a normalization  constant  given by
\begin{equation}
 c_n = \left[ \frac{(2L +n -1)(L+l+n-2)!}{2(L+l)!}\right]^{\frac{1}{2}}.
\end{equation}
Now we define scalar spherical harmonics in $n$ dimensions as
\begin{equation}
 Y_{Llp...m}(\chi, \theta,\phi \cdots) = {^n P^l_{L}}(\chi) Y_{lp...m}(\theta, \phi \cdots).
\end{equation}
Here $Y_{lpq....m}$ are the scalar spherical harmonics 
in $n$ dimensions and $Y_{ p q ....m}$ are the scalar spherical
 harmonics in $n-1$ dimensions.
Now the action of $\nabla^2$ on scalar spherical harmonic is 
\begin{equation}
 -\nabla^2 Y_{Llpq...m} = L(L+n -1)Y_{Llpq...m}.
\end{equation}
In this way we can construct the spherical harmonics from
 spherical harmonics on $S^1$.
Now we define the spherical harmonics on $S^1$ as 
 \begin{equation}
 Y_m=  \frac{1}{\sqrt{2\pi}}\exp(im\phi), 
 \end{equation}
so  we can construct spherical harmonics on $S^4$ as 
 \begin{equation}
  Y_{Llpm} = \frac{1}{\sqrt{2\pi}}c_4 c_2 c_3 \,\, {^4 P^l_L} {^3 P^p_l}
{^2 P^m_p} \exp(im\phi).
 \end{equation}
So for the scalar spherical harmonics on $S^4$, we have
\begin{equation}
 -\nabla^2 Y^{Llpm} = [L(L + 3) - 1] Y^{Llpm}.
\end{equation}
If $g$ is the metric on $S^4$, then $Y_{Llpm}$ are normalized as follows:
\begin{equation}
 \int d4x \sqrt{g} Y_{Llpm}Y^*_{L'l'p'm'} = \delta_{LL'}\delta_{ll'}\delta_{pp'}\delta_{mm'}.
\end{equation}
There are two kinds of vector spherical harmonics on $S^n$.
Let them be denoted by
$A_a^{Llp\cdots;0}$ and  $A_a^{Llp\cdots;1}$.
Then the action of $\nabla^2$ operator on them is given by
\begin{eqnarray}
  -\nabla^2 A_a^{Llp\cdots;0} &=& [L(L+n-1)-1] A_a^{Llp\cdots;0}, \nonumber \\
  -\nabla^2 A_a^{Llp\cdots;1} &=& [L(L+n-1)-1] A_a^{Llp\cdots;1}.
\end{eqnarray}
The divergence of vector spherical harmonics vanishes, so we have
\begin{eqnarray}
 \nabla^a A^{Llp\cdots;0}_a&=&0, \nonumber \\
 \nabla^a A^{Llpm;1}_a&=&0.
\end{eqnarray}
It can be shown that the vector spherical harmonics on $S^4$ are given by
\begin{eqnarray}
 A^{Llpm;1}_\chi &=& 0 , \\ A^{Llpm;1}_i& =& n_1 P_{L+1} A^{lpm}_i, \\
A^{Llpm; 0}_\chi &=& n_2 (\sin\chi)^{-2} P_{L+1}Y^{lpm},
 \\ A^{Llpm;0}_i &=& n_2  \frac{1}{\ell(\ell-2)} D_1 P_{L+1}\nabla_i Y^{lpm}. 
\end{eqnarray}
where  $Y^{lpm}$ and $A^{lpm}_i$ are scalar 
and vector spherical harmonics 
on $S^4$, respectively. The  
are normalization constants $n_1$ and $n_2$ are chosen, such that
\begin{equation}
 \int d^4 x \sqrt{g} g^{ab} A^{Llpm;0}_a A^{*Llpm;0}_b = 1,
\end{equation}
and 
\begin{equation}
 \int d^4 x \sqrt{g} g^{ab} A^{Llpm;1}_a A^{*Llpm;1}_b = 1.
\end{equation}
Now for the vector spherical harmonics on $S^4$, we have
\begin{eqnarray}
  -\nabla^2 A_a^{Llpm;0} &=& [L(L+3)-1] A_a^{Llpm;0}, \nonumber \\
  -\nabla^2 A_a^{Llpm;1} &=& [L(L+3)-1] A_a^{Llpm;1},
\end{eqnarray}
and 
\begin{eqnarray}
  \nabla^a A_a^{Llpm;0} &=& 0, \nonumber \\
  \nabla^a A_a^{Llpm;1} &=& 0.
\end{eqnarray}
\section{Faddeev-Popov for Yang-Mills Theories}
Now we mode expand a field for the Yang-Mills theory 
on $S^4 $ as
\begin{equation}
 T^A A^{A}_a =  \sum a^{Llpm, n} A_a^{Llmp;n},
\end{equation}
If we take the gauge fixing condition as the Lorentz gauge
\begin{equation}
 F[A] = \nabla_a A^{a} = 0,
\end{equation}
then the  gauge fixing condition for each of the modes is 
\begin{equation}
 F[A] = \nabla^a A_a^{ Llmp;n} = 0,
\end{equation}
where $n =0,1$.
Now term $\mathcal{L}_g$  is given by  
\begin{equation}
 \mathcal{L}_g = - \sqrt{g}[\frac{1}{2\alpha} 
\nabla_a A^{Aa} \nabla_b A^{Ab}],
\end{equation}
and the ghost field term is given by
\begin{equation}
 \mathcal{L}_{gh} = - i \sqrt{g}[\nabla^a \overline{c}^A D_a c^A].
\end{equation}
Here $D_a c^A$ is the covariant derivative and is given by
\begin{equation}
 D_a c^A = \nabla_a c^A + f^A_{BC} A_a^B c^C.
\end{equation}
Here $c^A$ and $\overline{c}^A$ are the ghosts and anti-ghosts 
respectively.
Now we can get the total Lagrangian as follows:
\begin{equation}
 \mathcal{L}_{t} = \mathcal{L} +  \mathcal{L}_g + \mathcal{L}_{gh}.
\end{equation}
We can  write the free part of the total action containing terms 
 quadratic in $A^{Aa}$ as follows:
\begin{equation}
S_{tf} = \int d^4x d^4x' \frac{\sqrt{g(x)}\sqrt{g(x')}}{2}A^{Aa}(x)
\mathcal{D}(x, x')_{ab'}(x,x')A^{Ab'}(x'),
 \end{equation}
where
\begin{equation}
\mathcal{D}(x,x')_{ab'} = \left(
  g_{ab}\nabla^2 + \frac{1-\alpha}{\alpha} \nabla_{b'} 
\nabla_a\right) \delta (x, x').
 \end{equation}
$\mathcal{D}(x,x')_{ab'}$ does not have any zero 
eigenvalue and so $\mathcal{D}(x,x')_{ab'}$ can be inverted.
 Its inverse is given by Green's function $ G (x, x')_{ab'}$,
 \begin{equation}
  G(x, x')_{ab'} = \mathcal{D}^{-1}(x,x')_{ab'}.\label{d}
 \end{equation}
Thus we have 
\begin{equation}
  G(x, x')_{ab'} =
 \sum N^{(L, \alpha)} A_a^{Llpm;n}(x)A_{b'}^{Llpm;n}(x').
\end{equation}
It may be noted that it is possible to choose a
 very special gauge, called axial gauge for 
Yang-Mills theories. In this gauge
 the ghosts decouple from the gauge fields. 
\section{BRST Symmetry}
We will now discuss the BRST for gauge theory on $S^4$. 
The sum of the gauge fixing Lagrangian and the ghost Lagrangian can be 
written as  
\begin{equation}
\mathcal{L}_{g} + \mathcal{L}_{gh} = \sqrt{g} s [- c \nabla_a A^{a} 
+ \frac{\alpha}{2} cb ].
\end{equation}
where  
$b$ does not contain any derivatives and the functional integral
 over $b$ can be done by completing the square. This way we will 
recover the original gauge fixing term.
The BRST transformations are give by 
\begin{eqnarray}
  s   {A}^A_a&=&  D_a c^A,\nonumber \\
 s   B^A &=& 0,\nonumber \\
 s   \overline{c}^A &=& B^A,\nonumber \\
 s   c^A &=& \frac{-i}{2}  f^{A}_{BC} c^B c^C.
\end{eqnarray}
Here the BRST transformation of ${A}_a ^A  $ is obtained by
 replacing the infinitesimal
 parameter $\Lambda^A$ in the gauge transformations by the
 ghost field $c^A$ and the BRST
 transformation of $c^A$ is obtained by taking 
the commutator of two gauge transformations and then replacing 
all the infinitesimal parameters by the $c^A$.
The BRST transformation of $B^A$ vanishes and
 BRST transformation of $\overline c^A$ is $B^A$. 
These BRST transformations are nilpotent
\begin{eqnarray}
 s^2 {A}^A_a&=& 0, \nonumber \\
 s^2 c^A &=& 0, \nonumber \\
 s^2 \overline c^A &=& 0, \nonumber \\
 s^2 B^A&=& 0.
\end{eqnarray}
Due to the nilpotency of these BRST transformations the sum 
of the gauge fixing and ghost terms is invariant under these BRST 
transformations, 
\begin{equation}
s\mathcal{L}_{g} + s\mathcal{L}_{gh} = \sqrt{g} s^2 [- c \nabla_a A^{a} 
+ \frac{\alpha}{2} cb ] =0.
\end{equation}
Now the original Lagrangian is invariant by itself as the action of 
$s$ on it just generates a gauge transformation 
with ghost fields acting as the gauge parameters
\begin{equation}
s \mathcal{L} = 0.
\end{equation}
Thus, the  total Lagrangian  $\mathcal{L}_t$ is invariant
 under  the BRST transformations.
\begin{equation}
 s\mathcal{L}_t =0.
\end{equation}

\section{Conclusion}
In this paper we have analysed the quantization of 
gauge theory on $S^4$. We have also analyzed the 
BRST symmetry of this model.
It is know that  gauge theories with BRST symmetry also have 
another symmetry called the anti-BRST symmetry. In anti-BRST symmetry
the role of the ghosts and anti-ghosts is interchanged. 
It will be interesting to analyse the anti-BRST symmetry of this model.

\end{document}